\documentclass[twocolumn,showpacs,preprintnumbers,amsmath,amssymb]{revtex4}

\usepackage{graphicx}
\usepackage{dcolumn}
\usepackage{bm}

\begin{document}

\noindent

\preprint{}

\title{Quantum fluctuations as deviation from classical dynamics of ensemble of trajectories parameterized by  unbiased hidden random variable}

\author{Agung Budiyono}

\affiliation{Institute for the Physical and Chemical Research, RIKEN, 2-1 Hirosawa, Wako-shi, Saitama 351-0198, Japan}

\date{\today}

\begin{abstract} 
  
A quantization method based on replacement of c-number by c-number parameterized by an unbiased hidden random variable is developed. In contrast to canonical quantization, the replacement has straightforward physical interpretation as statistical modification of classical dynamics of ensemble of trajectories, and implies a unique operator ordering. We then apply the method to develop quantum measurement without wave function collapse and external observer \`a la pilot-wave theory. 

\end{abstract}

\maketitle 

\section{Motivation}

The present paper discusses three closely interrelated aspects of quantum mechanics: canonical quantization of classical system, quantum-classical correspondence and measurement problem. The discussion will be confined to system of non-relativistic particles with no spin. It is well known that even in this case, despite the astonishing pragmatical successes of quantum mechanics, its foundation with regard to the above three aspects, is not without ambiguity \cite{Isham book}.  

Let us denote the position of the particles as $q=\{q_i\}$ and the corresponding conjugate momentum as $\underline{p}=\{\underline{p}_i\}$ where $i$ goes for all degree of freedom. In this paper all symbols with \underline{underline} is used to denote quantities satisfying the law of classical mechanics. In canonical quantization, given a classical quantity $F(q,\underline{p})$, then the so-called ``quantum observable'' is obtained by promoting $q$ and $\underline{p}$ to Hermitian operators $q\mapsto\hat{q}$ and $\underline{p}\mapsto\hat{p}$, and replacing the Poisson bracket by commutator $\{\cdot,\cdot\}\mapsto [\cdot,\cdot]/(i\hbar)$. This replacement of c-number (classical number) by q-number (quantum number/Hermitian operator) consequently does not in general give a unique Hermitian operator due to operator ordering ambiguity. An even deeper difficulty is lying at the conceptual level for while classical mechanics is developed using the basic notion of conventional trajectory, the resulting quantum mechanics does not refer to the notion of trajectory, except in the so-called pilot-wave interpretation \cite{Bohm paper} to be discussed later. A closely related question is that despite the fact that Planck constant plays a pivotal role in connecting the quantum and classical mechanics through quantization and classical limit, its physical origin is not clear. Elaborating this issue might open the way to discuss the limitation of quantum mechanics. 

On the other hand, in its standard interpretation, quantum mechanics is based on two different processes: unitary, continuous and deterministic evolution described by the Schr\"odinger equation when there is no measurement; and non-unitary, discontinuous and non-causal (random) process of wave function collapse in measurement \cite{measurement problem}. For general model of measurement, the first process alone will give a superposition of macroscopically distinct states, which in the standard interpretation leads to the so-called paradox of Schr\"odinger's cat. It is then assumed that measurement reveals, randomly, only one of the term in the superposition. Accordingly, the second process mentioned above is needed. Moreover, this interpretation assumes an apparatus which must behave according to classical mechanics. The whole system must then be divided into quantum system being measured and classical apparatus of measurement. It is however well-known that such line of division can be made anywhere, thus is ambiguous. Further, since in general quantum measurement does not reveal the pre-existing value prior measurement, then there is a question whether another apparatus is needed to probe the record of the first apparatus, which immediately leads to infinite regression. In the context of quantum-classical correspondence, one can thus ask why classical mechanics does not suffer from measurement problems mentioned above, and how the probability of finding in quantum measurement becomes the probability of being in classical measurement. 

Below we shall attempt to propose a solution to the above problems. Our basic idea is to first understand the physical meaning behind the formal rule of canonical quantization. We shall develop a quantization method by directly modifying classical dynamics of ensemble of trajectories parameterized by an unbiased hidden random variable. We shall show that given the classical Hamiltonians, the resulting equations for important class of physical systems can be rewritten into the Schr\"odinger equation with unique quantum Hamiltonians. The method is based on replacement of c-number by c-number, thus is free from operator ordering ambiguity. A couple examples where canonical quantization is ambiguous will be given. We shall further show that in all the cases to be considered, the particles posses effective velocity equal to the velocity of the particles in the pilot-wave theory and the Born's interpretation of wave function is valid for all time by construction. This allows us to describe quantum measurement without wave function collapse and external (classical) observer a la pilot-wave theory. However unlike the latter, our model is inherently stochastic and the wave function is not physically real.  

\section{Modification of classical dynamics of ensemble of trajectories using hidden random variable\label{modification of classical ensemble}}    

Let us consider the dynamics of $N$ particles system whose classical Hamiltonian is denoted by $\underline{H}(q,\underline{p};t)$. The classical dynamics of the particles is then given by the following Hamilton-Jacobi equation:
\begin{equation}
\partial_t\underline{S}(q;t)+\underline{H}(q,\partial_q\underline{S}(q;t);t)=0,
\label{H-J equation}
\end{equation}
where $\underline{S}(q;t)$ is the Hamilton principle function (HPF) so that the momentum field is given by $\underline{p}=\partial_q\underline{S}$ where $\partial_q=\{\partial_{q_i}\}$ \cite{Rund book}. Hamilton-Jacobi equation thus describes the dynamics of ensemble (congruence) of trajectories in configuration space. To solve this equation, one needs to set up an initial HPF $\underline{S}(q;0)$ which implies an initial classical momentum field $\underline{p}(q;0)=\partial_q\underline{S}(q;0)$. A single trajectory in configuration space is picked up if one also fixes the initial position of the particles.

Let us consider an ensemble of classical systems so that the position of the particles are initially  distributed in configuration space with probability density $\underline{\rho}(q;0)$, $\int dq\underline{\rho}(q;0)=1$. The probability density of the configuration of the particles at any time $\underline{\rho}(q;t)$ then satisfies the following continuity equation: 
\begin{equation}
\partial_t\underline{\rho}+\partial_{q}\cdot(\underline{v}(\underline{S})\underline{\rho})=0,
\label{continuity equation}
\end{equation}   
where $\underline{v}=\{{\underline{v}}_i\}$ is the classical velocity field. In the above equation, we have made explicit that in general, the classical velocity field $\underline{v}$ might depend on the HPF $\underline{S}$. Given a classical Hamiltonian, this relation can be obtained through (the Legendre transformation part of) the  Hamilton equation: 
\begin{equation}
{\underline{v}}_i=\frac{\partial \underline{H}}{\partial {\underline{p}}_i}\Big|_{\{{\underline{p}}_i=\partial_{q_i}\underline{S}\}}=f_i(\underline{S}),
\label{classical velocity field}
\end{equation}
where $f_i$, $i=1,\dots,N$ are some functions \cite{constrained motion}. The dynamics and statistics of the ensemble of classical trajectories are then obtained by solving Eqs. (\ref{H-J equation}), (\ref{continuity equation}) and (\ref{classical velocity field}) in term of  $\underline{S}(q;t)$, $\underline{\rho}(q;t)$ and $\underline{v}(q;t)$. 

Let us then proceed to develop a general scheme to modify the above dynamics of ensemble of classical trajectories. To do this, let us introduce a pair of real-valued functions, $S(q,\lambda;t)$ and $\Omega(q,\lambda;t)$, assumed to take over the role of $\underline{S}(q;t)$ and $\underline{\rho}(q;t)$, respectively, in the modified dynamics. Here $\lambda$ is a hidden random variable of action dimensional whose statistical properties will be specified later. Hence $\Omega(q,\lambda;t)$ is the joint-probability density that the particles are at configuration coordinate $q$ and the value of the hidden variable is $\lambda$. The marginal probability densities of the fluctuations of $q$ and $\lambda$ are thus given by 
\begin{eqnarray}
\rho(q;t)\doteq\int d\lambda\Omega(q,\lambda;t),\nonumber\\
P(\lambda)\doteq\int dq\Omega(q,\lambda;t),
\end{eqnarray}
where we have assumed that the statistics of $\lambda$ is independent of time.

Let us then propose the following rule of replacement to modify the classical dynamics of ensemble of trajectories governed by Eqs. (\ref{H-J equation}) and (\ref{continuity equation}):
\begin{eqnarray}
\underline{\rho}\mapsto\Omega,\hspace{25mm}\nonumber\\
\partial_{q_i}\underline{S}\mapsto\partial_{q_i}S+\frac{\lambda}{2}\frac{\partial_{q_i}\Omega}{\Omega}, \hspace {2mm}i=1,..,N,\nonumber\\
\partial_{t}\underline{S}\mapsto\partial_tS+\frac{\lambda}{2}\frac{\partial_t\Omega}{\Omega}+\frac{\lambda}{2}\partial_q\cdot f(S),\hspace{2mm}
\label{fundamental equation general}
\end{eqnarray}
where the functional form of $f=\{f_i\}$ is determined by the classical Hamiltonian according to Eq. (\ref{classical velocity field}). 

Let us first show that the replacement of Eq. (\ref{fundamental equation}) possesses a consistent classical correspondence if $S\rightarrow\underline{S}$ so that the Hamilton-Jacobi equation of (\ref{H-J equation}) is restored (notice that we have used the symbol ``$\mapsto$'' to denote replacement and ``$\rightarrow$'' to denote a limit). First, using the last two equations of (\ref{fundamental equation}), for sufficiently small $\Delta t$ and $\Delta q=\{\Delta q_i\}$, then expanding $\Delta F\doteq F(q+\Delta q;t+\Delta t)-F(q;t)\approx\partial_tF\Delta t+\partial_qF\cdot\Delta q$, for any function $F$, one has
\begin{equation}
\Delta\underline{S}\mapsto\Delta S+\frac{\lambda}{2}\Big(\frac{\Delta\Omega}{\Omega}+\partial_q\cdot f(S)\Delta t\Big). 
\label{statistical violation of HPSA}
\end{equation}  
One can see that in the limit $S\rightarrow\underline{S}$, in order to be consistent then the second term on the right hand side has to be vanishing. One thus has $d\Omega/dt=-\Omega\partial_q\cdot\underline{v}$, by Eq. (\ref{classical velocity field}). This is just the continuity equation of (\ref{continuity equation}). Hence, since $\underline{v}$ is independent of $\lambda$, in the limit $S\rightarrow\underline{S}$, one has $\rho=\int d\lambda\Omega\rightarrow\underline{\rho}$. We have thus a smooth classical correspondence.  

The next question is then what is the statistical properties of $\lambda$. We shall show in the next section that to reproduce the prediction of quantum mechanics, one needs to assume that the probability density function of $\lambda$ is given by 
\begin{equation}
P(\lambda)=\frac{1}{2}\delta(\lambda-\hbar)+\frac{1}{2}\delta(\lambda+\hbar), 
\label{Schroedinger condition}
\end{equation} 
where $\hbar$ is the reduced Planck constant. Namely, $\lambda$ can only take binary values $\lambda=\pm\hbar$ with equal probability.  

What we shall do in the following sections is as follows. First, given a classical Hamiltonian, we shall generate the classical dynamics of ensemble of trajectories according to Eqs. (\ref{H-J equation}), (\ref{continuity equation}) and (\ref{classical velocity field}). We then proceed to modify Eqs. (\ref{H-J equation}) and (\ref{continuity equation}) by imposing Eq. (\ref{fundamental equation general}). Averaging over the distribution of $\lambda$ and taking into account Eq. (\ref{Schroedinger condition}), we shall show that, for a class of important physical systems, the resulting equations can be put into the Schr\"odinger equation with a unique Hermitian quantum Hamiltonian. Below we shall assume that the fluctuations of $q$ and $\lambda$ are separable $\Omega(q,\lambda;t)=\rho(q;t)P(\lambda)$. Accordingly, Eq. (\ref{fundamental equation general}) becomes 
\begin{eqnarray}
\underline{\rho}\mapsto \rho P(\lambda),\hspace{25mm}\nonumber\\
\partial_{q_i}\underline{S}\mapsto\partial_{q_i}S+\frac{\lambda}{2}\frac{\partial_{q_i}\rho}{\rho}, \hspace {2mm}i=1,..,N,\nonumber\\
\partial_{t}\underline{S}\mapsto\partial_tS+\frac{\lambda}{2}\frac{\partial_t\rho}{\rho}+\frac{\lambda}{2}\partial_q\cdot f(S).\hspace{2mm}
\label{fundamental equation}
\end{eqnarray}

Let us note before proceeding that in the present paper we shall not discuss the issue of nonlocality \cite{Bell unspeakable}. For a review of the progress of hidden variable models in view of Bell nonlocality, see \cite{Genovese review HVM,Santos rebute}. Yet since we will claim that our model reproduces the prediction of quantum mechanics then it must violate Bell inequality. See however \cite{Nieuwenhuizen contextuality loophole} for an interesting discussion that the violation of Bell inequality does not necessarily lead to nonlocality due to the contextuality loophole. 

\section{Particle in external potential\label{particle in external potential}} 

\subsection{Emergent deterministic Schr\"odinger equation \label{emergent Schroedinger equation EM}}
 
Let us apply the above modification of classical mechanics to an ensemble of particle subjected to external potentials. For simplicity, let us consider the case of single particle with mass $m$. As will be seen, generalization to many particles is straightforward. The classical Hamiltonian is thus given by 
\begin{eqnarray}
\underline{H}=\frac{\big(\underline{p}-(e/c)A\big)^2}{2m}+eV, 
\label{classical Hamiltonian EM}
\end{eqnarray}
where $e$ is charge of the particle, $c$ is the velocity of light, $A(q;t)$ and $V(q;t)$ are the vector and scalar electromagnetic potentials, respectively. The Hamilton-Jacobi equation of (\ref{H-J equation}) thus reads
\begin{equation}
\partial_t\underline{S}+\frac{\big(\partial_q\underline{S}-(e/c)A\big)^2}{2m}+eV=0. 
\label{H-J equation EM}
\end{equation}
On the other hand, inserting Eq. (\ref{classical Hamiltonian EM}) into Eq. (\ref{classical velocity field}), the classical velocity field is related to $\underline{S}$ as 
\begin{equation}
\underline{v}=\big(\partial_q\underline{S}-(e/c)A\big)/m. 
\label{classical velocity field EM}
\end{equation}
The continuity equation of (\ref{continuity equation}) thus becomes
\begin{equation}
\partial_t\underline{\rho}+\frac{1}{m}\partial_q\cdot\Big(\big(\partial_q\underline{S}-(e/c)A\big)\underline{\rho}\Big)=0.
\label{continuity equation EM}
\end{equation}
Hence, the dynamics and statistics of classical ensemble of trajectories is governed by Eqs. (\ref{H-J equation EM}), (\ref{classical velocity field EM}) and (\ref{continuity equation EM}). 

Next, from Eq. (\ref{classical velocity field EM}) and the definition of $f$ given in Eq. (\ref{classical velocity field}), its functional form with respect to $\underline{S}$ is given by   
\begin{equation}
f(\underline{S})=(1/m)\big(\partial_q\underline{S}-(e/c)A\big). 
\label{magician EM}
\end{equation}
Equation (\ref{fundamental equation}) then becomes
\begin{eqnarray}
\underline{\rho}\mapsto\rho P(\lambda),\hspace{37mm}\nonumber\\
\partial_q\underline{S}\mapsto\partial_qS+\frac{\lambda}{2}\frac{\partial_q\rho}{\rho},\hspace{30mm}\nonumber\\
\partial_t\underline{S}\mapsto\partial_tS+\frac{\lambda}{2}\frac{\partial_t\rho}{\rho}+\frac{\lambda}{2m}\partial_q\cdot\big(\partial_qS-(e/c)A\big). 
\label{fundamental equation EM}
\end{eqnarray}

Let us investigate how the above set of equations modify Eqs. (\ref{H-J equation EM}) and (\ref{continuity equation EM}). First, imposing  the first two equations of (\ref{fundamental equation EM}) into Eq. (\ref{continuity equation EM}) one gets 
\begin{eqnarray}
\partial_t\rho+\frac{1}{m}\partial_q\cdot\Big(\rho\big(\partial_qS-(e/c)A\big)\Big)+\frac{\lambda}{2m}\partial_q^2\rho=0, 
\label{FPE EM}
\end{eqnarray}
where $\partial_q^2=\partial_q\cdot\partial_q$ and since $P(\lambda)$ is independent of time and space, it can be divided out. On the other hand, imposing the last two equations of (\ref{fundamental equation EM}) into Eq. (\ref{H-J equation EM}), one obtains
\begin{eqnarray}
\partial_tS+\frac{\big(\partial_qS-(e/c)A\big)^2}{2m}+eV-\frac{\lambda^2}{2m}\frac{\partial_q^2R}{R}\hspace{20mm}\nonumber\\
+\frac{\lambda}{2\rho}\Big(\partial_t\rho+\frac{1}{m}\partial_q\cdot\Big(\rho\big(\partial_qS-(e/c)A\big)\Big)+\frac{\lambda}{2m}\partial_q^2\rho\Big)=0,
\label{pre H-J-M for particle in electromagnetic field 1}
\end{eqnarray}
where we have defined $R\doteq\sqrt{\rho}$, and used the following identity:
\begin{equation}
\frac{1}{4}\frac{\partial_{q_i}\rho\partial_{q_j}\rho}{\rho^2}=\frac{1}{2}\frac{\partial_{q_i}\partial_{q_j}\rho}{\rho}-\frac{\partial_{q_i}\partial_{q_j}R}{R}, 
\label{fluctuation decomposition}
\end{equation} 
for the case of $i=j$. Inserting Eq. (\ref{FPE EM}) into Eq. (\ref{pre H-J-M for particle in electromagnetic field 1}), one has 
\begin{eqnarray}
\partial_tS+\frac{\big(\partial_qS-(e/c)A\big)^2}{2m}+eV-\frac{\lambda^2}{2m}\frac{\partial_q^2R}{R}=0.
\label{HJM EM}
\end{eqnarray}
We have thus pair of coupled equations (\ref{FPE EM}) and (\ref{HJM EM}) which still depends on the random variable $\lambda$. 

We shall proceed to take average of Eqs. (\ref{FPE EM}) and (\ref{HJM EM}) over the distribution of $\lambda$.  First, from Eq. (\ref{HJM EM}), since $R$ is independent of $\lambda$, one can see that $S(q,\lambda;t)$ and $S(q,-\lambda;t)$ satisfy the same differential equation. Assuming that initially $S$ possesses the following symmetry $S(q,\lambda;0)=S(q,-\lambda;0)$, the symmetry is then preserved for all the time
\begin{equation}
S(q,\lambda;t)=S(q,-\lambda;t). 
\label{phase symmetry}
\end{equation}
Let us now assume a specific form of $P(\lambda)$ given by Eq. (\ref{Schroedinger condition}): $
P(\lambda)=(1/2)\delta(\lambda-\hbar)+(1/2)\delta(\lambda+\hbar)$. Then averaging Eqs. (\ref{FPE EM}) and (\ref{HJM EM}) over the distribution of $\lambda$ and defining the following function:
\begin{eqnarray}
S_Q(q;t)\doteq\frac{1}{2}\Big(S(q,\hbar;t)+S(q,-\hbar;t)\Big)\nonumber\\
=S(q,\hbar;t)=S(q,-\hbar;t), 
\label{quantum phase}
\end{eqnarray}
which is valid by Eq. (\ref{phase symmetry}), one obtains the following pair of equations:
\begin{eqnarray}
\partial_t\rho+\frac{1}{m}\partial_q\cdot\Big(\rho\big(\partial_qS_Q-(e/c)A\big)\Big)=0,\hspace{3mm}\nonumber\\
\partial_tS_Q+\frac{\big(\partial_qS_Q-(e/c)A\big)^2}{2m}+eV-\frac{\hbar^2}{2m}\frac{\partial_q^2R}{R}=0.
\label{Madelung equation EM}
\end{eqnarray}

Let us further define the following complex-valued function:
\begin{equation}
\Psi_Q(q;t)\doteq R\exp(iS_Q/\hbar), 
\label{Schroedinger wave function}
\end{equation}
so that the probability density of the position of the particle is given by 
\begin{equation}
\rho(q;t)=|\Psi_Q(q;t)|^2. 
\label{Born's rule}
\end{equation}
The pair of equations in (\ref{Madelung equation EM}) can then be recast into  
\begin{equation}
i\hbar\partial_t\Psi_Q=\frac{1}{2m}\big(-i\hbar\partial_q-(e/c)A\big)^2\Psi_Q+eV\Psi_Q.
\label{Schroedinger equation EM}
\end{equation}
This is just the Schr\"odinger equation for a particle subjected to vector and scalar potentials $A(q;t)$ and $V(q;t)$ with the corresponding quantum Hamiltonian given by 
\begin{equation}
{\hat H}=\frac{1}{2m}\big(-i\hbar\partial_q-(e/c)A\big)^2+eV. 
\label{quantum Hamiltonian EM}
\end{equation}
One can also see from Eq. (\ref{Born's rule}) that the Born's interpretation of wave function is valid by construction.  

Further, from the lower equation of (\ref{Madelung equation EM}), it is straightforward to show that 
\begin{eqnarray}
\int dq d\lambda\rho(q;t)P(\lambda)(-\partial_tS)=\int dq\rho(q;t)(-\partial_tS_Q)\hspace{5mm}\nonumber\\
=\int dq\Psi_Q^*\Big(\frac{(-i\hbar\partial_q-(e/c)A)^2}{2m}+eV\Big)\Psi_Q. 
\label{quantum mechanical average energy}
\end{eqnarray}
The right hand side is just the quantum mechanical average energy which is conserved by the Schr\"odinger equation of (\ref{Schroedinger equation EM}). Hence, one should interpret $E\doteq-\partial_tS_Q$ as the effective energy of the particle.  

Note that while the Schr\"odinger equation is obtained when $\lambda$ can take only binary values $\pm\hbar$, $\lambda$ may be a function of a set of continuous random variables, say $\lambda=\lambda(\nu_1,\nu_2,\dots)$. For example, one may have $\lambda=\sqrt{\nu_1^2+\nu_2^2+\nu_3^2}=\pm\hbar$ where $\nu_i$, $i=1,2,3$, take continuous real value. One thus has to solve 
\begin{equation}
\nu_1^2+\nu_2^2+\nu_3^2=\hbar^2, 
\label{hidden ball}
\end{equation}
namely the point $\nu=\{\nu_1,\nu_2,\nu_3\}$ lies on the surface of a ball of radius $\hbar$. Now let us divide the surface of the ball into two with equal area and attribute to each division with $\pm$ signs. Then, if the point $\nu$ is distributed uniformly on the surface of the ball (say it moves sufficiently chaotic on the surface), the resulting $\lambda$ will satisfy Eq. (\ref{Schroedinger condition}).

The time-dependent Schr\"odinger equation is thus obtained through specific choice of $P(\lambda)$ given by Eq. (\ref{Schroedinger condition}). This result suggests that generalization of Schr\"odinger equation might be attained by allowing the probability density function of $\lambda$ to deviate from Eq. (\ref{Schroedinger condition}). This might further lead to possible correction to the prediction of quantum mechanics \cite{AgungDQM2}.

Let us mention here that there are many approaches reported in the literature to derive the Schr\"odinger equation with quantum Hamiltonian of the type given in Eq. (\ref{quantum Hamiltonian EM}) \cite{Feynman PI,Nelson stochastic mechanics,de la Pena stochastic mechanics,Santamato,Frieden-Fisher,Nagasawa transformation,Garbaczewski-Vigier,Nottale,Reginatto-Fisher,Olavo,Kaniadakis,Fritsche,Markopoulou,Parwani: information measure,Smolin,Santos.SE,Brenig invariant uncertainty principle,Groessing,Rusov,Gog,de la Pena SED}. The advantage of our derivation is three folds. First, it is derived in the scheme of quantizing a general classical Hamiltonian. Hence, taking aside that the solution exists, the method can be applied directly to other class of classical Hamiltonians. Second, it is derived by modifying the classical dynamics of ensemble of trajectories so that the quantum-classical correspondence is conceptually kept transparent. In particular, we have no problem of conceptual jump in quantum-classical transition from a quantum theory which does not refer to conventional notion of trajectories to a classical theory which is founded based on the notion of trajectories. The classical limit of Schr\"odinger equation is given by the dynamics of ensemble of classical trajectories. Finally, the Schr\"odinger equation is shown to correspond to  a specific distribution of hidden random variable. Hence, it hints to a straightforward generalization \cite{AgungDQM2}.  

Let us note before proceeding that the hidden variable $\lambda$ is not the property of a single particle. Rather, it suggests the existence of background field which pervades all space whose detail interaction with the particle is not known, resulting in the stochastic motion of the particle. The presence of background field is also assumed in Nelson's stochastic mechanics \cite{Nelson stochastic mechanics} and stochastic electrodynamics \cite{de la Pena SED} approaches to explain the origin of quantum fluctuations. Next, as shown above, to get the correct time evolution, we have to first calculate the solutions (in the form of differential equation parameterized by the hidden variable) and then take average over the distribution of the hidden variable. The converse will lead to wrong time evolution. The situation is more like random walk. Namely, one has to evolve the walker (the particle) using the random step to obtain the correct time evolution, and then do the averaging over the probability of each single step. Taking the average of the single step in random walk will lead to trivial motion. In this case, to be meaningful, the fluctuations of the hidden variable $\lambda$ has to be much faster than the fluctuations of $q$. 

\subsection{Effective velocity and pilot-wave theory}

First, the upper equation of (\ref{Madelung equation EM}) can be regarded as a generalized continuity equation so that one can read off an effective velocity field of the ensemble of particle which is given by
\begin{equation}
v(S_Q)\doteq\frac{1}{m}\big(\partial_qS_Q-(e/c)A\big)=f(S_Q), 
\label{effective velocity EM}  
\end{equation}
where in the last equality we have used Eq. (\ref{magician EM}). On the other hand, if $\lambda$ satisfies Eq. (\ref{Schroedinger condition}), then as shown in Eq. (\ref{Schroedinger wave function}), $S_Q$ is just the phase of Schr\"odinger wave function $\Psi_Q$. Hence, in this case, the numerical value of the effective velocity of the particles is equal to the actual velocity of the particle in pilot-wave theory \cite{Bohm paper}. One can also see from Eq. (\ref{effective velocity EM}) that $\partial_qS_Q$ should be interpreted as the effective momentum field of ensemble of the particle. We have thus an effectively similar picture with pilot-wave theory in the sense that  the particle always possesses definite position and momentum and it moves ``as if'' it is guided by the wave function so that the effective velocity is given by Eq. (\ref{effective velocity EM}).  

Our model however differs from pilot-wave theory as follows. The latter is based on the assumption that: (a) for any dynamical system, the Schr\"odinger equation and the corresponding guidance relation are postulated; (b) the time evolution is deterministic; (c) the wave function $\Psi_Q(q;t)$ is physically real field; and (d) the initial distribution of the particle is assumed to be given by $\rho(q;0)=|\Psi_Q(q;0)|^2$  to reproduce the prediction of quantum mechanics. In particular, the last two assumptions constitute one of its main critics \cite{Pauli-criticsm,Keller-criticsm}. With the assumption that the wave function is physically real field, first, it is not clear how to prepare an ensemble to satisfy $\rho(q;0)=|\Psi_Q(q;0)|^2$. See however Refs. \cite{Bohm-Vigier,Valentini H-theorem,Duerr-Goldstein-Zanghi} for argumentation against this critics. Second, why there is no back-reaction from the particle to the wave function like for example in the particle-field interaction in the theory of  electromagnetic. Unlike pilot-wave theory, in our model, the (effective) deterministic time evolution governed by the Schr\"odinger equation and the corresponding guidance relation emerge naturally from a statistical modification of classical dynamics rather than postulated. The original dynamics is inherently stochastic. Moreover, the wave function is not physically real. It is just a mathematical tool to describe the dynamics and statistics of the ensemble of trajectories, and $\rho(q;t)=|\Psi_Q(q;t)|^2$ is valid for all time by construction. Nevertheless, despite the above conceptual difference, one can conclude that our model will reproduce the pilot-wave theory prediction on statistical wave-like pattern in slits experiment and tunneling of potential barrier \cite{interference and tunneling in PWT}. 

\subsection{Quantization: unique ordering and quantum-classical correspondence\label{quantization of classical Hamiltonian}}

We have mentioned in the previous subsection that the scheme to derive the Schr\"odinger equation presented in subsection \ref{emergent Schroedinger equation EM} can be viewed as to provide a method of quantization of  classical Hamiltonian. In fact, the quantum Hamiltonian of Eq. (\ref{quantum Hamiltonian EM}) can be obtained from the classical Hamiltonian of Eq. (\ref{classical Hamiltonian EM}) by replacing the classical momentum with the quantum momentum operator 
\begin{equation}
\underline{p}\mapsto{\hat p}\doteq-i\hbar\partial_q,
\end{equation}
as prescribed by the canonical quantization in configuration space representation (see however the discussion at the end of the present subsection). Hence, for the case of particle in external potentials, our method correctly reproduces the result of canonical quantization.    
 
As mentioned in Section I, however, given a classical Hamiltonian, the canonical quantization rule in general will give an infinite alternatives of quantum Hamiltonians due to the operator ordering ambiguity. In contrast to this, it is evident that the method of quantization proposed in the present paper will lead to unique Hermitian quantum Hamiltonian, if the solution exists. This came from the fact that while in canonical quantization one replaces c-number by q-number, in our method, c-number is replaced by another c-number parameterized by random hidden variable as prescribed by Eq. (\ref{fundamental equation}). To give an example where the canonical quantization rule leads to ambiguity, let us consider the dynamics of particle with position-dependent mass $m(q)$ which has wide applications in solid state physics \cite{Lakshmanan,Galbraith,Young,Einvoll,Arias,Leblond,Serra,Yu,Plastino,Carinena,Kuru,Bagchi,Sever,Zhang,Nieto,Cruz,Kerimov,Jana,Levai}. For simplicity let us assume that the particle is free. The classical Hamiltonian then takes the form  
\begin{equation}
\underline{H}=B(q)\underline{p}^2,
\label{classical Hamiltonian position-dependent mass}
\end{equation}
where $B(q)=1/(2m)$ is a real-valued differentiable function of $q$. Using canonical quantization, one can then choose one out of infinite alternatives of quantum Hamiltonians. For example, if $B(q)\sim q^2$ up to some multiplicative constant, then one can either choose for the corresponding Hermitian quantum Hamiltonian $({\hat p}^2{\hat q}^2+{\hat q}^2{\hat p}^2)/2$ or ${\hat p}{\hat q}^2{\hat p}$ which are related to each other, by virtue of the canonical commutation relation $[\hat{q},\hat{p}]=i\hbar$, as ${\hat p}{\hat q}^2{\hat p}=({\hat p}^2{\hat q}^2+{\hat q}^2{\hat p}^2)/2+\hbar^2$. 

Let us show that our method of quantization leads to unique Hermitian quantum Hamiltonian with a specific ordering of operators. First, given the classical Hamiltonian of Eq. (\ref{classical Hamiltonian position-dependent mass}), the Hamilton-Jacobi equation of (\ref{H-J equation}) reads
\begin{equation}
\partial_t\underline{S}+B(\partial_q\underline{S})^2=0. 
\label{H-J equation position-dependent mass}
\end{equation}
Substituting Eq. (\ref{classical Hamiltonian position-dependent mass}) into Eq. (\ref{classical velocity field}), the classical velocity field is given by 
\begin{equation}
\underline{v}=2B\partial_q\underline{S}. 
\label{classical velocity position-dependent mass}
\end{equation}
Hence, the continuity equation of (\ref{continuity equation}) becomes
\begin{equation}
\partial_t\underline{\rho}+2\partial_q\cdot\big(B\underline{\rho}\partial_q\underline{S}\big)=0. 
\label{continuity equation position-dependent mass}
\end{equation}
Next, from Eq. (\ref{classical velocity position-dependent mass}), $f$ defined in Eq. (\ref{classical velocity field}) is given by $f(\underline{S})=2B\partial_q\underline{S}$ so that Eq. (\ref{fundamental equation}) becomes 
\begin{eqnarray}
\underline{\rho}\mapsto\rho P(\lambda),\hspace{23mm}\nonumber\\
\partial_q\underline{S}\mapsto\partial_qS+\frac{\lambda}{2}\frac{\partial_q\rho}{\rho},\hspace{15mm}\nonumber\\
\partial_t\underline{S}\mapsto\partial_tS+\frac{\lambda}{2}\frac{\partial_t\rho}{\rho}+\lambda\partial_q\cdot\big(B\partial_qS\big).
\label{fundamental equation position-dependent mass}
\end{eqnarray}

Let us see how the above set of equations modify the pair of Eqs. (\ref{H-J equation position-dependent mass}) and (\ref{continuity equation position-dependent mass}). Imposing the first two equations of (\ref{fundamental equation position-dependent mass}) into Eq. (\ref{continuity equation position-dependent mass}) one gets
\begin{equation}
\partial_t\rho+2\partial_q(B\rho\partial_qS)+\lambda\partial_q(B\partial_q\rho)=0. 
\label{FPE position-dependent mass}
\end{equation}
On the other hand, imposing the last two equations of (\ref{fundamental equation position-dependent mass}) into Eq. (\ref{H-J equation position-dependent mass}), one obtains 
\begin{eqnarray}
\partial_tS+B(\partial_qS)^2-\lambda^2\Big(B\frac{\partial_q^2R}{R}+\partial_qB\frac{\partial_q R}{R}\Big)\nonumber\\
+\frac{\lambda}{2\rho}\Big(\partial_t\rho+2\partial_q(B\rho\partial_qS)+\lambda\partial_q(B\partial_q\rho)\Big)=0,
\label{u position-dependent mass}
\end{eqnarray}
where we have again defined $R\doteq\sqrt{\rho}$ and used the identity of Eq. (\ref{fluctuation decomposition}) for $i=j$. Substituting Eq. (\ref{FPE position-dependent mass}) into Eq. (\ref{u position-dependent mass}) one  gets 
\begin{equation}
\partial_tS+B(\partial_qS)^2-\lambda^2\Big(B\frac{\partial_q^2R}{R}+\partial_qB\frac{\partial_q R}{R}\Big)=0.
\label{HJM position-dependent mass}
\end{equation}
We have thus pair of coupled equations (\ref{FPE position-dependent mass}) and (\ref{HJM position-dependent mass}) which still depend on the hidden variable $\lambda$. 

One can then see that $S(q,\hbar;t)=S(q,-\hbar;t)=S_Q(q;t)$ satisfies the same equation (\ref{HJM position-dependent mass}) where $\lambda^2$ is replaced by $\hbar^2$. Hence, averaging over the fluctuations of the parameter $\lambda=\pm\hbar$ which is assumed to be equally probable, Eqs. (\ref{FPE position-dependent mass}) and (\ref{HJM position-dependent mass}) become
\begin{eqnarray}
\partial_t\rho+2\partial_q(B\rho\partial_qS_Q)=0,\hspace{20mm}\nonumber\\
\partial_tS_Q+B(\partial_qS_Q)^2-\hbar^2\Big(B\frac{\partial_q^2R}{R}+\partial_qB\frac{\partial_q R}{R}\Big)=0.
\end{eqnarray}

Finally, recalling Eq. (\ref{Schroedinger wave function}) that  $\Psi_Q=R\exp(iS_Q/\hbar)$, the above pair of equations can be recast into 
\begin{equation}
i\hbar\partial_t\Psi_Q=-\frac{\hbar^2}{2}\Big(B\partial_q^2+\partial_q^2B\Big)\Psi_Q
+\frac{\hbar^2}{2}(\partial_q^2B)\Psi_Q. 
\end{equation}
This is just the Schr\"odinger equation with quantum Hamiltonian given by ${\hat H}={\hat p}B(q){\hat p}$. One thus obtains the following quantization mapping
\begin{equation}
B(q)\underline{p}^2\mapsto{\hat p}B(q){\hat p}. 
\label{quantization position-dependent mass}
\end{equation}
In particular for constant $B$, one has 
\begin{equation}
\underline{p}^2\mapsto\hat{p}^2. 
\label{quantum Hamiltonian kinetic energy}
\end{equation}

Let us mention that the same result as in Eq. (\ref{quantization position-dependent mass}) is also reported in the derivation of Schr\"odinger equation using the principle of exact-uncertainty and a principle of extremization of ensemble of Hamiltonian density \cite{Hall ordering}. However, in contrast to the latter which can only be applied to quantize a specific type of classical Hamiltonian with a quadratic momentum dependence, our method formally applies as well, as will be shown below and in the next section, to classical Hamiltonian which contains a term that is linear in momentum. One might then expect that Eq. (\ref{quantum Hamiltonian kinetic energy}) can be extended to any power in momentum, namely $\underline{p}^n\mapsto\hat{p}^n$, where $n$ is integer. The answer is however negative. A straightforward calculation to quantize a classical Hamiltonian which is proportional to $\underline{p}^3$, regardless of its physical meaning, will not lead to a Schr\"odinger equation with quantum Hamiltonian proportional to $\hat{p}^3$. 

Next, let us assume that the  classical Hamiltonian under interest is decomposable as $\underline{H}=a\underline{H}_1+b\underline{H}_2$, where $a$ and $b$ are real numbers. Then from Hamilton equation, the classical velocity field is also decomposable into $\underline{v}(\underline{S})=(\partial\underline{H}/\partial\underline{p})|_{\{\underline{p}=\partial_q\underline{S}\}}=af_1(\underline{S})+bf_2(\underline{S})$, where the function $f_i$ corresponds to $\underline{H}_i$, $i=1,2$. Hence, $f$ defined in Eq. (\ref{classical velocity field}) is also decomposable into $f(\underline{S})=af_1(\underline{S})+bf_2(\underline{S})$. Further let us assume that each term of the decomposition of classical Hamiltonian is mapped into a quantum Hamiltonian as $\underline{H}_1\mapsto{\hat H}_1$ and $\underline{H}_2\mapsto{\hat H}_2$.  Keeping all of these in mind and recalling that $\underline{H}$ and $\underline{v}$ appear linearly in Eqs. (\ref{H-J equation}) and (\ref{continuity equation}) and also the linearity of the Schr\"odinger equation, one can conclude that applying the quantization method to the total classical Hamiltonian one will get \begin{equation}
\underline{H}=a\underline{H}_1+b\underline{H}_2\mapsto{\hat  H}= a{\hat H}_1+b{\hat H}_2. 
\label{linearity} 
\end{equation}
The quantization mapping induced by our hidden variable model is thus linear. 

To apply the above property, let us first formally quantize the following classical Hamiltonian
\begin{equation}
\underline{H}=B(q)\underline{p},
\label{classical Hamiltonian linear momentum}
\end{equation}
which is assumed to be one of the term of a physically sensible Hamiltonian, say a term that appears in Eq. (\ref{classical Hamiltonian EM}). Here $B(q)$ is a differentiable function of $q$. Applying the method of quantization developed in this paper one formally has the following quantization mapping:
\begin{equation}
B(q)\underline{p}\mapsto\frac{B{\hat p}+{\hat p}B}{2}. 
\label{quantum Hamiltonian linear momentum}
\end{equation} 
Detail of the calculation is given in Appendix \ref{quantization of linear momentum}. In particular, for  $B(q)=1$ one has $\underline{p}\mapsto\hat{p}=-i\hbar\partial_q$ and putting $\underline{p}=1$, one has $B(q)\mapsto B(q)$, or formally $q\mapsto\hat{q}=q$. 

Let us apply the above results to quantize the classical Hamiltonian of a particle in electromagnetic field given in Eq. (\ref{classical Hamiltonian EM}). First, Eq. (\ref{classical Hamiltonian EM}) can be expanded into
\begin{equation}
\underline{H}=\frac{\underline{p}^2}{2m}-\frac{eA\underline{p}}{mc}+\frac{e^2A^2}{2mc^2}+eV. 
\label{classical Hamiltonian EM expanded}
\end{equation}
Applying the quantization mapping of Eqs. (\ref{quantum Hamiltonian kinetic energy}) and (\ref{quantum Hamiltonian linear momentum}), recalling the linearity of the quantization mapping of Eq. (\ref{linearity}), one then obtains 
\begin{eqnarray}
{\hat H}=\frac{\hat{p}^2}{2m}-\frac{e}{2mc}\big(A\hat{p}+\hat{p}A\big)+\frac{e^2A^2}{2mc^2}+eV, \nonumber
\label{quantum Hamiltonian EM expanded}
\end{eqnarray}
which is equal to Eq. (\ref{quantum Hamiltonian EM}), as expected. Hence, in developing Eq. (\ref{quantum Hamiltonian EM}) from Eq. (\ref{classical Hamiltonian EM}) using canonical quantization by directly promoting the classical momentum into quantum momentum operator $\underline{p}\mapsto\hat{p}$, one is implicitly assuming the ordering given in Eq. (\ref{quantum Hamiltonian linear momentum}). In contrast to this, in our hidden variable model for quantization, Eq. (\ref{quantum Hamiltonian linear momentum}) is derived rather than assumed. 

\section{Measurement of momentum, position, angular momentum without wave function collapse\label{measurement}}

In the present section, we shall apply the modification of classical dynamics developed in the previous section using the type of hidden random variable with probability density given by Eq. (\ref{Schroedinger condition}) to a class of classical model of measurement of momentum, position and angular momentum. 

\subsection{Classical measurement}

Let us first discuss a class of measurement model in classical mechanics.  Let us consider the dynamics of two interacting particles, the first particle with coordinate $q_1$ represents the system to be measured and the other with coordinate $q_2$ represents the measuring apparatus. Let us suppose that one wants to measure a physical quantity $\underline{A}_1$ of the system. It is a function of the position $q_1$ and classical momentum ${\underline{p}}_1$, $\underline{A}_1=\underline{A}_1(q_1,{\underline{p}}_1)$. To do this, let us choose the following classical measurement-interaction Hamiltonian: 
\begin{equation}
\underline{H}=g\underline{A}_1(q_1,{\underline{p}}_1){\underline{p}}_2, 
\label{classical Hamiltonian with measurement-interaction}
\end{equation}
where $g$ is a coupling constant. Let us further assume that the interaction is impulsive so that the individual free Hamiltonians of the particles are ignorable. 

$\underline{A}_1$ is thus conserved $d\underline{A}_1/dt=\{\underline{A}_1,\underline{H}\}=0$. The idea is then to correlate the value of $\underline{A}_1(q_1,{\underline{p}}_1)$ with the classical momentum of the apparatus ${\underline{p}}_2$ while keeping the value of $\underline{A}_1(q_1,{\underline{p}}_1)$ unchanged. On the other hand, one also has $dq_2/dt=\{q_2,\underline{H}\}=g\underline{A}_1$, which can be integrated to give $q_2(T)=q_2(0)+g\underline{A}_1T$, where $T$ is time span of the measurement-interaction. The value of $\underline{A}_1$ prior to the measurement can thus be inferred from the observation of the initial and final values of $q_2$ (the pointer of the apparatus). Since each measurement reveals the value of $A_1$ prior-measurement then there is no need to introduce a second apparatus (third particle) to observe the position of the second particle (the pointer of the first apparatus).  

Below we shall modify the classical dynamics of ensemble of trajectories generated by classical Hamiltonian of Eq. (\ref{classical Hamiltonian with measurement-interaction}) for measurement of momentum, position and angular momentum following the method developed in the previous section. Momentum and position represent physical quantities with continuous quantum mechanical spectrum. They are also important in view of canonical commutation relation and Heisenberg uncertainty relation. On the other hand, angular momentum represents physical quantity with discrete quantum mechanical spectrum. The crucial problem in this model is that whether one needs further ``quantum apparatus'' to observe $q_2(t)$, the position of the apparatus pointer. We shall show that this is not the case. Namely the model with interacting two particles will be shown to be sufficient for this purpose. 

We have to mention that in reality, however, the above model with one dimensional apparatus is oversimplified. To this end let us emphasize that the aim of the discussion is only to show that in principle, the method of quantization proposed in the present paper can lead to measurement without wave function collapse and necessitating no external (classical) observer. In this respect, we believe that if it does not work for one degree of freedom then it will be more difficult to expect that it will work for realistic measurement model. Especially, our model excludes the irreversibility of the registration process which can only be done by realistic apparatus plus bath using large degree of freedom. See Ref. \cite{Theo} for an elaborated discussion of quantum measurement using realistic model of apparatus. 

\subsection{Quantum Hamiltonian for the measurement of momentum, position and angular momentum\label{quantum Hamiltonian with measurement-interaction}}

\subsubsection{Quantum Hamiltonian for the measurement of momentum}

Let us first discuss the case of momentum measurement. One thus puts $\underline{A}_1={\underline{p}}_1$ into Eq. (\ref{classical Hamiltonian with measurement-interaction}) to have the following measurement-interaction classical Hamiltonian:
\begin{equation}
\underline{H}_p=g{\underline{p}}_1{\underline{p}}_2.
\label{classical Hamiltonian of momentum}
\end{equation} 
In impulsive measurement, the Hamilton-Jacobi equation of (\ref{H-J equation}) then reads
\begin{equation}
\partial_t\underline{S}+g\partial_{q_1}\underline{S}\partial_{q_2}\underline{S}=0.
\label{H-J equation momentum}
\end{equation}
On the other hand, inserting Eq. (\ref{classical Hamiltonian of momentum}) into Eq. (\ref{classical velocity field}), one obtains the following classical velocity field for the two particles
\begin{eqnarray}
{\underline{v}}_1=g\partial_{q_2}\underline{S},\hspace{3mm}{\underline{v}}_2=g\partial_{q_1}\underline{S}. 
\label{velocity field for momentum}
\end{eqnarray}
The continuity equation of (\ref{continuity equation}) thus becomes
\begin{equation}
\partial_t\underline{\rho}+g\partial_{q_1}(\underline{\rho}\partial_{q_2}\underline{S})+g\partial_{q_2}(\underline{\rho}\partial_{q_1}\underline{S})=0.
\label{continuity equation momentum} 
\end{equation}
From Eq. (\ref{velocity field for momentum}), $f$ defined in Eq. (\ref{classical velocity field}) takes the form 
\begin{eqnarray}
f_1(\underline{S})=g\partial_{q_2}\underline{S},\hspace{3mm}f_2(\underline{S})=g\partial_{q_1}\underline{S}, 
\label{actual velocity field momentum}
\end{eqnarray}
so that $\partial_q\cdot f(S)=2g\partial_{q_1}\partial_{q_2}S$. Hence, Eq. (\ref{fundamental equation}) becomes
\begin{eqnarray}
\underline{\rho}\mapsto\rho P(\lambda),\hspace{23mm}\nonumber\\
\partial_{q_i}\underline{S}\mapsto\partial_{q_i}S+\frac{\lambda}{2}\frac{\partial_{q_i}\rho}{\rho},\hspace{2mm}i=1,2,\nonumber\\
\partial_{t}\underline{S}\mapsto\partial_tS+\frac{\lambda}{2}\frac{\partial_t\rho}{\rho}+ g\lambda\partial_{q_1}\partial_{q_2}S.
\label{fundamental equation momentum}
\end{eqnarray} 

Let us proceed to investigate the change brought by Eq. (\ref{fundamental equation momentum}) onto the Hamilton-Jacobi equation of (\ref{H-J equation momentum}) and the corresponding continuity equation of (\ref{continuity equation momentum}) describing ensembles of classical trajectories. First, imposing the first two equations of (\ref{fundamental equation momentum}) into Eq. (\ref{continuity equation momentum}) one obtains 
\begin{equation}
\partial_t\rho+g\partial_{q_1}(\rho\partial_{q_2}S)+g\partial_{q_2}(\rho\partial_{q_1}S)+g\lambda\partial_{q_1}\partial_{q_2}\rho=0.
\label{FPE momentum} 
\end{equation}
On the other hand, imposing the last two equations of (\ref{fundamental equation momentum}) into Eq. (\ref{H-J equation momentum}) one gets 
\begin{eqnarray}
\partial_tS+g\partial_{q_1}S\partial_{q_2}S-g\lambda^2\frac{\partial_{q_1}\partial_{q_2}R}{R}
+\frac{\lambda}{2\rho}\Big(\partial_t\rho\nonumber\\
+g\partial_{q_1}(\rho\partial_{q_2}S)+g\partial_{q_2}(\rho\partial_{q_1}S)+g\lambda\partial_{q_1}\partial_{q_2}\rho\Big)=0,
\label{uu m}
\end{eqnarray}
where we have defined $R\doteq\sqrt{\rho}$ and used Eq. (\ref{fluctuation decomposition}). Substituting Eq. (\ref{FPE momentum}) into Eq. (\ref{uu m}) one thus has 
\begin{equation}
\partial_tS+g\partial_{q_1}S\partial_{q_2}S-g\lambda^2\frac{\partial_{q_1}\partial_{q_2}R}{R}=0.
\label{HJM equation momentum}
\end{equation} 
We have thus pair of coupled equations (\ref{FPE momentum}) and (\ref{HJM equation momentum}) which still depend on the random variable $\lambda$ whose probability density function is assumed to be given by Eq. (\ref{Schroedinger condition}). 

One can then see that $S(q,\hbar;t)=S(q,-\hbar;t)=S_Q(q;t)$ satisfies the same differential equation of (\ref{HJM equation momentum}) with $\lambda^2$ replaced by $\hbar^2$. Then averaging Eqs. (\ref{FPE momentum}) and (\ref{HJM equation momentum}) over the distribution of $\lambda=\pm\hbar$ with equal probability one obtains the following pair of equations:
\begin{eqnarray}
\partial_t\rho+g\partial_{q_1}(\rho\partial_{q_2}S_Q)+g\partial_{q_2}(\rho\partial_{q_1}S_Q)=0,\nonumber\\
\partial_tS_Q+g\partial_{q_1}S_Q\partial_{q_2}S_Q-g\hbar^2\frac{\partial_{q_1}\partial_{q_2}R}{R}=0.
\label{Madelung equation momentum} 
\end{eqnarray}
Finally, recalling Eq. (\ref{Schroedinger wave function}) that $\Psi_Q=\sqrt{\rho}\exp(iS_Q/\hbar)=R\exp(iS_Q/\hbar)$, the above pair of equations can be combined into the Schr\"odinger equation $i\hbar\partial_t\Psi_Q={\hat H}_p\Psi_Q$ with measurement-interaction quantum Hamiltonian 
\begin{equation}
{\hat H}_p=g{\hat p}_1{\hat p}_2,
\label{quantum Hamiltonian momentum} 
\end{equation}
where ${\hat p}_i=-i\hbar\partial_{q_i}$, $i=1,2$. Again, by construction one has $\rho(q;t)=|\Psi_Q(q;t)|^2$. Moreover, from the upper equation of (\ref{Madelung equation momentum}), the effective velocity is $f(S_Q)$ where $f$ is given by Eq. (\ref{actual velocity field momentum}) so that it is equal to the actual velocity field of the particles in pilot-wave theory.  

\subsubsection{Quantum Hamiltonian for the measurement of position \label{subsubsection measurement of position}}

Next let us consider the measurement of position. One thus put $\underline{A}_1=q_1$ into Eq. (\ref{classical Hamiltonian with measurement-interaction}) to have the following classical measurement-interaction Hamiltonian: 
\begin{equation}
\underline{H}_q=gq_1{\underline{p}}_2.
\label{interaction Hamiltonian position}
\end{equation}
The Hamilton-Jacobi equation of (\ref{H-J equation}) thus reads
\begin{equation}
\partial_t\underline{S}+gq_1\partial_{q_2}\underline{S}=0. 
\label{H-J equation position}
\end{equation}
On the other hand, inserting Eq. (\ref{interaction Hamiltonian position}) into Eq. (\ref{classical velocity field}), the classical velocity field is given by 
\begin{equation}
{\underline{v}}_1=0,\hspace{2mm}{\underline{v}}_2=gq_1.
\label{classical velocity field position}
\end{equation}
The above pair of equations provide constraint to the dynamics of the particles. Hence, the continuity equation of (\ref{continuity equation}) becomes 
\begin{equation}
\partial_t\underline{\rho}+gq_1\partial_{q_2}\underline{\rho}=0. 
\label{continuity equation position}
\end{equation}
From Eq. (\ref{classical velocity field position}) and the definition of $f$ given in Eq. (\ref{classical velocity field}) one has
\begin{equation}
f_1=0,\hspace{2mm} f_2=gq_1,
\label{actual velocity field position} 
\end{equation}
so that $\partial_q\cdot f=0$. Equation (\ref{fundamental equation}) thus becomes
\begin{eqnarray}
\underline{\rho}\mapsto\rho P(\lambda),\hspace{19mm}\nonumber\\
\partial_{q_i}\underline{S}\mapsto\partial_{q_i}S+\frac{\lambda}{2}\frac{\partial_{q_i}\rho}{\rho},\hspace{2mm}i=1,2,\nonumber\\
\partial_{t}\underline{S}\mapsto\partial_tS+\frac{\lambda}{2}\frac{\partial_t\rho}{\rho}.\hspace{10mm}
\label{fundamental equation position}
\end{eqnarray}

Now let us apply the above set of equations to Eqs. (\ref{H-J equation position}) and (\ref{continuity equation position}). First, imposing the first equation of (\ref{fundamental equation position}) into Eq. (\ref{continuity equation position}) one obtains 
\begin{equation}
\partial_t\rho+gq_1\partial_{q_2}\rho=0, 
\label{continuity equation position: quantum}
\end{equation}
namely Eq. (\ref{continuity equation position}) is kept unchanged. Next, imposing the last two equations of (\ref{fundamental equation position}) into Eq. (\ref{H-J equation position}) one obtains 
\begin{equation}
\partial_tS+gq_1\partial_{q_2}S+\frac{\lambda}{2\rho}(\partial_t\rho+gq_1\partial_{q_2}\rho)=0. 
\end{equation}
Substituting Eq. (\ref{continuity equation position: quantum}) one thus gets 
\begin{equation}
\partial_tS+gq_1\partial_{q_2}S=0. 
\label{HJM position}
\end{equation}
Again, Eq. (\ref{H-J equation position}) remains unchanged. We have thus pair of decoupled equations (\ref{continuity equation position: quantum}) and (\ref{HJM position}). 

Notice then that $\lambda$ does not appear explicitly as parameter. Identifying $S_Q=S$, and defining $\Psi_Q\doteq\sqrt{\rho}\exp(iS_Q/\hbar)$ so that $|\Psi_Q(q;t)|^2=\rho(q;t)$, the pair of Eqs. (\ref{continuity equation position: quantum}) and (\ref{HJM position}) can then be combined together into the Schr\"odinger equation $i\hbar\partial_t\Psi_Q={\hat H}_q\Psi_Q$ with quantum Hamiltonian
\begin{equation}
{\hat H}_q=gq_1{\hat p}_2.
\label{quantum Hamiltonian position}
\end{equation}
Again, one can see from Eq. (\ref{continuity equation position: quantum}) that the effective velocity of the particles is $f(S_Q)$ where $f$ is given by Eq. (\ref{actual velocity field position}). Hence, it is again equal to the velocity of the particles in pilot-wave theory.   

\subsubsection{Quantum Hamiltonian for the measurement of angular momentum}

Let us proceed to develop the quantum Hamiltonian for the measurement of angular momentum. To make explicit the three dimensional nature of the problem, let us put $q_1=(x_1,y_1,z_1)$. For simplicity let us first consider the measurement of $z-$part of angular momentum. In this case $\underline{A}_1$ in Eq. (\ref{classical Hamiltonian with measurement-interaction}) takes the form $\underline{A}_1=x_1{\underline{p}}_{y_1}-y_1{\underline{p}}_{x_1}$, where ${\underline{p}}_{x_1}$ is the conjugate momentum of $x_1$ and so on, so that the measurement-interaction classical Hamiltonian of Eq. (\ref{classical Hamiltonian with measurement-interaction}) reads
\begin{equation}
\underline{H}_l=g(x_1{\underline{p}}_{y_1}-y_1{\underline{p}}_{x_1}){\underline{p}}_2. 
\label{classical Hamiltonian angular momentum}
\end{equation}
The Hamilton-Jacobi equation of (\ref{H-J equation}) thus becomes
\begin{equation}
\partial_t\underline{S}+g\big(x_1\partial_{y_1}\underline{S}-y_1\partial_{x_1}\underline{S}\big)\partial_{q_2}\underline{S}=0. 
\label{H-J equation angular momentum}
\end{equation}

On the other hand, substituting Eq. (\ref{classical Hamiltonian angular momentum}) into Eq. (\ref{classical velocity field}), the classical velocity field is given by 
\begin{eqnarray}
{\underline{v}}_{x_1}=-gy_1\partial_{q_2}\underline{S},\hspace{2mm}{\underline{v}}_{y_1}=gx_1\partial_{q_2}\underline{S},\hspace{2mm}{\underline{v}}_{z_1}=0,\nonumber\\
{\underline{v}}_2=g\big(x_1\partial_{y_1}\underline{S}-y_1\partial_{x_1}\underline{S}\big). \hspace{15mm}
\label{classical velocity angular momentum}
\end{eqnarray} 
Hence, the continuity equation of (\ref{continuity equation}) becomes
\begin{eqnarray}
\partial_t\underline{\rho}-gy_1\partial_{x_1}(\underline{\rho}\partial_{q_2}\underline{S})+gx_1\partial_{y_1}(\underline{\rho}\partial_{q_2}\underline{S})\nonumber\\
+gx_1\partial_{q_2}(\underline{\rho}\partial_{y_1}\underline{S})-gy_1\partial_{q_2}(\underline{\rho}\partial_{x_1}\underline{S})=0. 
\label{continuity equation angular momentum}
\end{eqnarray}

Next, from Eq. (\ref{classical velocity angular momentum}), $f$ defined in Eq. (\ref{classical velocity field}) takes the form  
\begin{eqnarray}
f_{x_1}(\underline{S})=-gy_1\partial_{q_2}\underline{S},\hspace{2mm}f_{y_1}(\underline{S})=gx_1\partial_{q_2}\underline{S},\hspace{2mm}f_{z_1}(\underline{S})=0,\nonumber\\
f_2(\underline{S})=g\big(x_1\partial_{y_1}\underline{S}-y_1\partial_{x_1}\underline{S}\big). \hspace{15mm}
\label{actual velocity field angular momentum}
\end{eqnarray}
One thus has $\partial_q\cdot f(S)=2g(x_1\partial_{q_2}\partial_{y_1}S-y_1\partial_{q_2}\partial_{x_1}S)$. Substituting this into Eq. (\ref{fundamental equation}), one then obtains 
\begin{eqnarray}
\underline{\rho}\mapsto\rho P(\lambda),\hspace{30mm}\nonumber\\
\partial_{x_1}\underline{S}\mapsto\partial_{x_1}S+\frac{\lambda}{2}\frac{\partial_{x_1}\rho}{\rho},\hspace{20mm}\nonumber\\
\partial_{y_1}\underline{S}\mapsto\partial_{y_1}S+\frac{\lambda}{2}\frac{\partial_{y_1}\rho}{\rho},\hspace{20mm}\nonumber\\
\partial_{q_2}\underline{S}\mapsto\partial_{q_2}S+\frac{\lambda}{2}\frac{\partial_{q_2}\rho}{\rho},\hspace{20mm}\nonumber\\
\partial_{t}\underline{S}\mapsto\partial_{t}S+\frac{\lambda}{2}\frac{\partial_{t}\rho}{\rho}+ g\lambda(x_1\partial_{y_1}\partial_{q_2}S-y_1\partial_{x_1}\partial_{q_2}S).\hspace{0mm}
\label{fundamental equation angular momentum}
\end{eqnarray}

Let us proceed to see how the above set of equations modify Eqs. (\ref{H-J equation angular momentum}) and (\ref{continuity equation angular momentum}). Imposing the first four equations of (\ref{fundamental equation angular momentum}) into Eq. (\ref{continuity equation angular momentum}) one obtains, after a simple calculation
\begin{eqnarray}
\partial_t\rho-gy_1\partial_{x_1}(\rho\partial_{q_2}S)+gx_1\partial_{y_1}(\rho\partial_{q_2}S)+gx_1\partial_{q_2}(\rho\partial_{y_1}S)\nonumber\\
-gy_1\partial_{q_2}(\rho\partial_{x_1}S)-g\lambda(y_1\partial_{x_1}\partial_{q_2}\rho-x_1\partial_{y_1}\partial_{q_2}\rho)=0.\nonumber\\
\label{FPE angular momentum}
\end{eqnarray}
On the other hand, imposing the last four equations of (\ref{fundamental equation angular momentum}) into Eq. (\ref{H-J equation angular momentum}), one has, after an arrangement
\begin{eqnarray}
\partial_tS+g\big(x_1\partial_{y_1}S-y_1\partial_{x_1}S\big)\partial_{q_2}S-g\lambda^2\Big(x_1\frac{\partial_{y_1}\partial_{q_2}R}{R}\nonumber\\-y_1\frac{\partial_{x_1}\partial_{q_2}R}{R}\Big)+\frac{\lambda}{2\rho}\Big(\partial_t\rho-gy_1\partial_{x_1}(\rho\partial_{q_2}S)\nonumber\\
+gx_1\partial_{y_1}(\rho\partial_{q_2}S)+gx_1\partial_{q_2}(\rho\partial_{y_1}S)-gy_1\partial_{q_2}(\rho\partial_{x_1}S)\nonumber\\
-g\lambda(y_1\partial_{x_1}\partial_{q_2}\rho-x_1\partial_{y_1}\partial_{q_2}\rho)\Big)=0,
\label{ccc}
\end{eqnarray}
where $R\doteq\sqrt{\rho}$ and we have used Eq. (\ref{fluctuation decomposition}). Substituting Eq. (\ref{FPE angular momentum}), the last term in the bracket vanishes to give 
\begin{eqnarray}
\partial_tS+g\big(x_1\partial_{y_1}S-y_1\partial_{x_1}S\big)\partial_{q_2}S\hspace{20mm}\nonumber\\
-g\lambda^2\Big(x_1\frac{\partial_{y_1}\partial_{q_2}R}{R}-y_1\frac{\partial_{x_1}\partial_{q_2}R}{R}\Big)=0.
\label{HJM angular momentum}
\end{eqnarray}
One thus has pair of coupled equations (\ref{FPE angular momentum}) and (\ref{HJM angular momentum}) which are parameterized by the random variable $\lambda$. 

Again, one can see that $S(q,\hbar;t)=S(q,-\hbar;t)=S_Q(q;t)$ satisfies the same differential equation of (\ref{HJM angular momentum}) where $\lambda^2$ is replaced by $\hbar^2$. Hence, taking the average of Eqs. (\ref{FPE angular momentum}) and (\ref{HJM angular momentum}) over the distribution of $\lambda=\pm\hbar$ with equal probability as in Eq. (\ref{Schroedinger condition}) gives the following pair of equations:
\begin{eqnarray}
\partial_t\rho-gy_1\partial_{x_1}(\rho\partial_{q_2}S_Q)+gx_1\partial_{y_1}(\rho\partial_{q_2}S_Q)\hspace{10mm}\nonumber\\
+gx_1\partial_{q_2}(\rho\partial_{y_1}S_Q)-gy_1\partial_{q_2}(\rho\partial_{x_1}S_Q)=0.\nonumber\\
\partial_tS+g\big(x_1\partial_{y_1}S_Q-y_1\partial_{x_1}S_Q\big)\partial_{q_2}S_Q\hspace{20mm}\nonumber\\
-g\hbar^2\Big(x_1\frac{\partial_{y_1}\partial_{q_2}R}{R}-y_1\frac{\partial_{x_1}\partial_{q_2}R}{R}\Big)=0.
\label{Madelung equation for angular momentum measurement}
\end{eqnarray}

Finally, recalling Eq. (\ref{Schroedinger wave function}) that $\Psi_Q=R\exp(iS_Q/\hbar)$ so that $|\Psi_Q(q;t)|^2=R(q;t)^2=\rho(q;t)$, the above pair of equations can be recast into the Schr\"odinger equation $i\hbar\partial_t\Psi_Q={\hat H}_l\Psi_Q$ with quantum Hamiltonian 
\begin{equation}
{\hat H}_l=g{\hat L}_{z_1}{\hat p}_2,
\label{Hamiltonian operator angular momentum}
\end{equation} 
where ${\hat L}_{z_1}=x_1{\hat p}_{y_1}-y_1{\hat p}_{x_1}=-i\hbar (x_1\partial_{y_1}-y_1\partial_{x_1})$. As expected, ${\hat L}_{z_1}$ is just the $z-$component of the (quantum mechanical) angular momentum operator.  Moreover, one can again see from the upper equation of (\ref{Madelung equation for angular momentum measurement}) that the effective velocity of the particles are $f(S_Q)$ where $f$ is given by Eq. (\ref{actual velocity field angular momentum}) so that it is equal to the actual velocity of the particles in pilot-wave theory. The above results can be extended to measurement of angular momentum along $x-$ and $y-$ directions straightforwardly using cyclic permutation among the coordinates $(x_1,y_1,z_1)$. One will then obtain the Schr\"odinger equation with quantum Hamiltonian of Eq. (\ref{Hamiltonian operator angular momentum}) where $\hat{L}_{z_1}$ is replaced by the quantum mechanical angular momentum operators along the $x-$ and $y-$ directions, respectively. Moreover, since in principle one can take any direction as $z-$axis, then the above result applies as well to angular momentum measurement along any direction.   

\subsection{Measurement without wave function collapse and external observer}

In the previous section, starting from a class of classical Hamiltonian for the measurement of momentum, position and angular momentum, $\underline{H}=g\underline{A}_1\underline{p}_2$, where $\underline{A}_1$ is the physical quantities being measured, we have arrived at the following Schr\"odinger equation:
\begin{equation}
i\hbar\partial_t\Psi_Q={\hat H}\Psi_Q=g{\hat A}_{1}{\hat p}_2\Psi_Q,
\label{Schroedinger equation measurement-interaction}
\end{equation} 
where $\hat{A}_1$ is a Hermitian operator given by ${\hat A}_1=\underline{A}_1(q,{\hat p}_1)$. Our hidden variable model of quantization thus reproduces the results of canonical quantization. However, unlike the latter, in all of the cases considered, one can identify an effective velocity of the particles which turns out to be equal to the actual velocity of the particles in pilot-wave theory, and the Born's interpretation of wave function, $|\Psi_Q(q;t)|^2=\rho(q;t)$, is valid for all time, by construction. We can then follow all the argumentation of the pilot-wave theory in describing the process of measurement without wave function collapse \cite{Bohm paper}. 

To do this, let us first assume that ${\hat A}_1$ has discrete spectrum as the case for measurement of angular momentum. Hence, one has ${\hat A}_1\psi_n(q_1)=a_n\psi_n(q_1)$, $n=0,1,2,\dots$, where $a_n$ is the real-valued eigenvalue of ${\hat A}_1$ and $\psi_n$ is the corresponding eigenfunction. $\{\psi_n\}$ thus makes a complete set of orthonormal functions. Then, ignoring the free Hamiltonian for impulsive measurement-interaction, the Schr\"odinger equation of (\ref{Schroedinger equation measurement-interaction}) has the following general solution: 
\begin{equation}
\Psi_Q(q_1,q_2;t)=\sum_nc_n\psi_n(q_1)\varphi(q_2-ga_nt), 
\label{entanglement system-apparatus}
\end{equation}
where $\varphi(q_2)$ is the initial wave function of the apparatus which is assumed to be sufficiently localized, $c_n$ is complex number, and $\phi(q_1)\doteq\sum_nc_n\psi_n(q_1)$ is the initial wave function of the system. In other words, $c_n$ is the coefficient of expansion of the initial wave function of the system in term of the orthonormal set of the eigenfunctions of ${\hat A}_1$. 

For sufficiently large $g$, $\varphi_n(q_2)\doteq\varphi(q_2-ga_nT)$ is not overlapping for different $n$ and each is correlated to a distinct $\psi_n(q_1)$. One then argues, following pilot-wave theory \cite{Bohm paper}, that the outcome of single measurement event corresponds to the packet $\varphi_n(q_2)$ which is actually entered by the apparatus particle. Namely, if $q_2(t)$ belongs to the support of $\varphi_n(q_2)$, then we admit that the result of measurement is given by $a_n$. This can be generalized to ${\hat A}_1$ with continuous spectrum, as the case for the measurement of momentum or position, by replacing the sum in Eq. (\ref{entanglement system-apparatus}) with integration. As in pilot-wave theory, the probability to find the outcome $a_n$ is given by $P(a_n)=|c_n|^2$, that is the experimentally well-verified Born's statistical rule. This can be shown as direct implication of $\rho(q;t)=|\Psi_Q(q;t)|^2$. The prediction of quantum mechanics is thus reproduced without invoking wave function collapse induced by external (classical) observer. 

Notice that the linearity of the Schr\"odinger equation plays a very pivotal role in the discussion of measurement. The superposition of solution in Eq. (\ref{entanglement system-apparatus}) is made possible by the linearity of the Schr\"odinger equation of (\ref{Schroedinger equation measurement-interaction}). Since $\varphi_n(q_2)=\varphi(q_2-ga_nT)$ refers to the wave function of pointer of the apparatus, then it has been argued within the standard quantum mechanics that Eq. (\ref{entanglement system-apparatus}) is a superposition of macroscopically distinct states. This leads to the paradox of Schr\"odinger's cat suggesting an indefiniteness of the state of macroscopic body which is against our everyday experience (recall that in the standard quantum mechanics, an observable possesses definite value only when the state is an eigenfunction of the observable which is not the case for Eq (\ref{entanglement system-apparatus})). It is to save this situation that in the standard quantum mechanics one needs to invoke a wave function collapse to get one of the term in the superposition of Eq. (\ref{entanglement system-apparatus}) \cite{Ballentine paper}. 

This paradox however is based on the assumption that the superposition of states of Eq. (\ref{entanglement system-apparatus}) refers to an individual system (and apparatus) and that the description of an individual system by the wave function is complete \cite{Ballentine paper}. In contrast to this, in our dynamical model, the superposition of state, or in general any wave function, describes an ensemble of identically prepared system rather than individual system. Moreover, the description of an individual system by the wave function is not complete: a single system is always described by definite values of position and momentum of the particles and an unbiased random variable $\lambda=\pm\hbar$. In this respect, the superposition of state in Eq. (\ref{entanglement system-apparatus}) does not mean macroscopic indefiniteness since at any time, the pointer always possesses definite position. Hence, there is no paradox of Schr\"odinger's cat and accordingly there is no need to invoke the wave function collapse to get one term of the superposition as required by the standard quantum mechanics.  

Further, one can see in the discussion of the previous subsection that the measurement of position is different from the measurement of momentum and angular momentum. Namely, unlike in the two latter cases, in the case of position measurement, Eq. (\ref{fundamental equation position}) does not change the classical Hamilton-Jacobi and continuity equations of (\ref{H-J equation position}) and (\ref{continuity equation position}). Both pair of functions $(\underline{S},\underline{\rho})$ and $(S,\rho)$ satisfy the same pair of equations, that of Eqs. (\ref{H-J equation position}) and (\ref{continuity equation position}). Hence, the classical results of measurement is preserved by Eq. (\ref{fundamental equation position}): there is no quantum correction. Conversely, the Schr\"odinger equation with quantum Hamiltonian of Eq. (\ref{quantum Hamiltonian position}) can be rewritten into the classical Hamilton-Jacobi equation of (\ref{H-J equation position}) and the continuity equation of (\ref{continuity equation position}) describing classical dynamics of ensemble of trajectories. One can thus conclude that, as in the case of classical measurement, it is possible to reveal the pre-existing value of the position immediately prior to the measurement. On the other hand, for the cases of measurement of momentum and angular momentum, the results of the measurement are not equal to the pre-existing values possessed by the systems. In this regards, the measurement of position is special. The derivation of the quantum Hamiltonian of measurement of position also shows that the ability to write the dynamics of ensemble of trajectories into the Schr\"odinger equation is not sufficient to distinguish quantum from classical mechanics. 

The above results on position measurement further leads to an important implication. Recall that the results of the measurement of momentum and angular momentum are inferred from the position of the second particle (apparatus pointer). Then one might argue that one needs another, the third particle, as the second apparatus to probe the position of the second particle (the first apparatus). Proceeding in this way thus will lead to infinite regression: one will further need the forth particle (the third apparatus) to probe the position of the third particle (the second apparatus) and so on. In our model, however, since the quantum treatment of the position measurement is equivalent to the classical treatment revealing the position of the particle prior-measurement, then the second measurement on the position of the second particle (the first apparatus) is in principle not necessary. Namely, the results of position measurement by the second, third, forth apparatuses and so on are all equal to each other. 

\subsection{Quantum mechanical observable and quantum-classical correspondence \label{quantization of physical quantity}}

First, the development of quantum Hamiltonian with measurement-interaction ${\hat H}=g{\hat A}_1{\hat p}_2$ from the corresponding classical Hamiltonian $\underline{H}=g\underline{A}_1(q_1,\underline{p}_1)\underline{p}_2$ in subsection \ref{quantum Hamiltonian with measurement-interaction} can be formally summarized into the following mapping
\begin{eqnarray}
\underline{p}_2\mapsto\hat{p}_2,\hspace{2mm}\underline{A}_1\mapsto\hat{A}_1. 
\end{eqnarray}
Hence, it can be regarded as the quantization of classical quantity $\underline{A}_1$ to get the corresponding Hermitian operator $\hat{A}_1$ in the context of measurement. $\hat{A}_1$ is called as ``quantum observable'' in the standard formalism of quantum mechanics. As shown in subsection \ref{quantum Hamiltonian with measurement-interaction}, for the case where $\underline{A}_1$ is momentum, position and angular momentum, the corresponding Hermitian operator ${\hat A}_1$ can be obtained formally by the following substitution rule: $\underline{p}_1\mapsto\hat{p}_1= -i\hbar\partial_{q_1}$ and $q_1\mapsto{\hat q}_1=q_1$. For these specific but fundamental dynamical variables, our method thus reproduces the results of canonical quantization. In contrast to the latter, however, the quantization method reported in the present paper is developed by directly modifying classical dynamics of ensemble of measurement parameterized by an unbiased binary random variable $\lambda=\pm\hbar$. We have thus a continuous and transparent transition from quantum to classical measurement. 

Further, recall that $[{\hat q}_i,{\hat p}_j]=i\hbar\delta_{ij}$ leads to the Heisenberg uncertainty relation $\sigma_{q_i}\sigma_{p_i}\ge\hbar/2$, where $\sigma_{q_i}$ and $\sigma_{p_i}$ are the standard deviation of results of measurement of position and the corresponding conjugate momentum in ensemble of identically prepared systems. Our dynamical model thus shows that the Heisenberg uncertainty relation is a direct implication of modification of classical dynamics for ensemble of trajectories as prescribed by Eq. (\ref{fundamental equation}) being applied to measurement. In particular, in the limit where $S\rightarrow\underline{S}$, one smoothly regains the classical dynamics so that $\sigma_{q_i}\sigma_{p_i}\ge 0$.    

An immediate question then arises whether the method of quantization of classical quantity in the context of measurement developed in the present paper can be applied to any classical quantities, namely any function of position and classical momentum $F=F(q,\underline{p})$. To discuss this matter, first, it is not clear even in the classical mechanics whether any arbitrary function $F(q,\underline{p})$ is physically meaningful at all. In reality, hitherto, for spin-less particle, only position, momentum, angular momentum and energy have unambiguous physical meaning. Second, even if $F(q,\underline{p})$ is physically meaningful, it is not clear whether it can be measured directly. This is due to the fact that in reality measurement is done by mapping the properties of the system being measured to non-overlapping subsets of the configuration space. Hence, measurement-interaction is a special type of interaction. This gives a physical limitation to the kind of classical quantities which can be directly measured. 

Taking all the above physical aspects aside, in contrast to canonical quantization which in general leads to infinite alternative of Hermitian operators for a given general classical quantity which is the direct implication of replacing c-number by q-number, it is evident that the method of quantization in the context of measurement model with classical Hamiltonian of Eq. (\ref{classical Hamiltonian with measurement-interaction}) presented in this paper, which is based on replacement of c-number by c-number, will give a unique Hermitian observable, if a solution exists. An example of the quantization of classical quantity of the type $B(q)\underline{p}$ in the context of measurement, where canonical quantization leads to ambiguity, is given in appendix \ref{quantization of classical quantity}. 

\section{Conclusion and discussion}

We have proposed a quantization method by modifying the classical dynamics of ensemble of trajectories. The deviation from the classical mechanics is characterized by pair of real-valued functions $S(q,\lambda;t)$ and $\Omega(q,\lambda;t)$ parameterized by a hidden random variable $\lambda$ with specific statistical property following the rule of Eq. (\ref{fundamental equation}). In the classical limit, $S(q,\lambda;t)$ and $\Omega(q,\lambda;t)$ reduce into the Hamilton principle function $\underline{S}(q;t)$ and the classical probability density of the position $\underline{\rho}(q;t)$. Given a classical Hamiltonian, the model is applied to system of particles in external potentials, with position-dependent mass, and to a class of classical measurement of momentum, position and angular momentum. We showed that the resulting equations can be put into the Schr\"odinger equation with unique Hermitian quantum Hamiltonian. The wave function refers to ensemble of system rather than to an individual system. In contrast to the canonical quantization which replaces c-number by q-number implying operator ordering ambiguity, our method is based on replacement of c-number by c-number, thus is free from the problem of operator ordering ambiguity. The canonical commutation relation $[{\hat q}_i,{\hat p}_j]=i\hbar\delta_{ij}$, which lies at the bottom of the canonical quantization, is thus given statistical and dynamical meaning as a  modification of classical dynamics of ensemble of trajectories in configuration space parameterized by an unbiased hidden random variable. This offers a conceptually smooth and physically transparent quantum-classical correspondence. 

We then identified an effective velocity of the particles which turns out to be equal to the velocity of the particles in pilot-wave theory. However, unlike pilot-wave theory, our model is strictly stochastic, the wave function is not physically real and the Born's interpretation of wave function is valid by construction. This allows us to conclude that our model will reproduce the prediction of pilot-wave theory on statistical wave-like pattern in single and double slits experiments and also in tunneling of potential barrier. Moreover, following the argumentation of pilot-wave theory, we then developed the process of measurement without wave function collapse and external observer, reproducing the statistical prediction of quantum mechanics. Since our dynamical model of measurement reduces into the classical dynamics of measurement when $S\rightarrow\underline{S}$, one can conclude that in this limit, the probability of finding of quantum measurement reduces into the probability of being of classical measurement. In this sense, we have thus argued that quantum mechanics is an emergence statistical phenomena \cite{Adler}.  

A common pragmatical question against any alternative approaches to quantum mechanics is that whether it offers new testable predictions which can not be calculated using the standard formalism of quantum mechanics. This is a very hard wall to tunnel in view of the pragmatical successes of the quantum mechanics. In our approach, however, since the Schr\"odinger equation is shown to be emergent corresponding to a specific type of distribution of hidden random variable $P(\lambda)$ given by Eq. (\ref{Schroedinger condition}), then we may expect that it will lead to new prediction if $P(\lambda)$ is allowed to deviate from Eq. (\ref{Schroedinger condition}). This, for example can be attained by allowing $|\lambda|$ to fluctuate around $\hbar$ with very small yet finite width. We shall elaborate the detail implications of this idea in separate work \cite{AgungDQM2}.    

\begin{acknowledgments} 

\end{acknowledgments}

\appendix

\section{\label{quantization of linear momentum}}

Let us quantize a classical Hamiltonian which takes the following form:
\begin{equation}
\underline{H}=B(q)\underline{p}, 
\label{classical Hamiltonian linear momentum}
\end{equation}
which is assumed to be part of a physically sensible Hamiltonian, and $B(q)$ is a differentiable function of $q$. First, the Hamilton-Jacobi equation of (\ref{H-J equation}) becomes
\begin{equation}
\partial_t\underline{S}+B\partial_q\underline{S}=0. 
\label{H-J equation linear momentum}
\end{equation} 
Further, inserting Eq. (\ref{classical Hamiltonian linear momentum}) into Eq. (\ref{classical velocity field}), the classical velocity field is given by 
\begin{equation}
\underline{v}=B. 
\label{classical velocity linear momentum}
\end{equation}
This provides a constraint to the motion of the particle. Thus, the continuity equation of (\ref{continuity equation}) reads
\begin{equation}
\partial_t\underline{\rho}+\partial_q(B\underline{\rho})=0. 
\label{continuity equation linear momentum}
\end{equation}

Next, from Eq. (\ref{classical velocity linear momentum}), $f$ defined in Eq. (\ref{classical velocity field}) is given by $f=B$, so that Eq. (\ref{fundamental equation}) becomes
\begin{eqnarray}
\underline{\rho}\mapsto\rho P(\lambda),\hspace{18mm}\nonumber\\
\partial_q\underline{S}\mapsto\partial_qS+\frac{\lambda}{2}\frac{\partial_q\rho}{\rho},\hspace{8mm}\nonumber\\
\partial_t\underline{S}\mapsto\partial_tS+\frac{\lambda}{2}\frac{\partial_t\rho}{\rho}+\frac{\lambda}{2}\partial_qB.
\label{fundamental equation linear momentum}
\end{eqnarray}

Now let us apply the above set of equations to modify Eqs. (\ref{H-J equation linear momentum}) and (\ref{continuity equation linear momentum}). First, imposing the upper equation of  (\ref{fundamental equation linear momentum}), Eq. (\ref{continuity equation linear momentum}) becomes
\begin{equation}
\partial_t\rho+\partial_q(B\rho)=0. 
\label{continuity equation linear momentum: quantum}
\end{equation}
Hence, Eq. (\ref{continuity equation linear momentum}) is kept unchanged. Further, imposing the last two equations of (\ref{fundamental equation linear momentum}) into Eq. (\ref{H-J equation linear momentum}) one obtains 
\begin{equation}
\partial_tS+B\partial_qS+\frac{\lambda}{2\rho}(\partial_t\rho+\partial_q(B\rho))=0. 
\end{equation}
Inserting Eq. (\ref{continuity equation linear momentum: quantum}) one thus has
\begin{equation}
\partial_tS+B\partial_qS=0. 
\label{HJM linear momentum}
\end{equation}
Namely, Eq. (\ref{H-J equation linear momentum}) is also kept unchanged. We have thus pair of decoupled equations (\ref{continuity equation linear momentum: quantum}) and (\ref{HJM linear momentum}). 

Notice then that $\lambda$ does not appear explicitly as a parameter of the resulting differential equations. Identifying $S_Q=S$, and defining $\Psi_Q\doteq\sqrt{\rho}\exp(iS_Q/\hbar)$ so that $|\Psi_Q(q;t)|^2=\rho(q;t)$, the pair of Eqs. (\ref{continuity equation linear momentum: quantum}) and (\ref{HJM linear momentum}) can then be combined together into the following Schr\"odinger equation: 
\begin{equation}
i\hbar\partial_t\Psi_Q=-i\frac{\hbar}{2}(B\partial_q+\partial_qB)\Psi_Q, 
\label{Schroedinger equation linear momentum}
\end{equation}
from which one can extract a Hermitian quantum Hamiltonian as 
\begin{equation}
{\hat H}=\frac{B(-i\hbar\partial_q)+(-i\hbar\partial_q)B}{2}=\frac{B{\hat p}+{\hat p}B}{2}. 
\end{equation}

\section{\label{quantization of classical quantity}}

Let us quantize the classical quantity of the type $F=B(q)\underline{p}$  in the context of measurement discussed in Section \ref{measurement}, where $B(q)$ is a differentiable function of $q$. One thus put $A_1=B(q_1)\underline{p}_1$ into Eq. (\ref{classical Hamiltonian with measurement-interaction}) so that the classical Hamiltonian for the interaction-measurement is given by 
\begin{equation}
\underline{H}=gB(q_1)\underline{p}_1\underline{p}_2. 
\label{classical Hamiltonian classical quantity}
\end{equation}
Notice that $B$ does not depend on $q_2$, the coordinate of the apparatus. The Hamilton-Jacobi equation of (\ref{H-J equation}) thus reads
\begin{equation}
\partial_t\underline{S}+gB\partial_{q_1}\underline{S}\partial_{q_2}\underline{S}=0. 
\label{H-J equation classical quantity}
\end{equation}
Next, inserting Eq. (\ref{classical Hamiltonian classical quantity}) into Eq. (\ref{classical velocity field}) one has 
\begin{equation}
\underline{v}_1=gB\partial_{q_2}\underline{S},\hspace{2mm}\underline{v}_2=gB\partial_{q_1}\underline{S}. 
\label{classical velocity field classical quantity}
\end{equation}
The continuity equation of (\ref{continuity equation}) thus becomes 
\begin{equation}
\partial_t\underline{\rho}+g\partial_{q_1}\big(\underline{\rho} B\partial_{q_2}\underline{S}\big)+g\partial_{q_2}\big(\underline{\rho}B\partial_{q_1}\underline{S}\big)=0. 
\label{continuity equation classical quantity}
\end{equation}

On the other hand, from Eq. (\ref{classical velocity field classical quantity}), $f$ defined in Eq. (\ref{classical velocity field}) is given by 
\begin{equation}
f_1(\underline{S})=gB\partial_{q_2}\underline{S},\hspace{2mm}f_2(\underline{S})=gB\partial_{q_1}\underline{S}. 
\label{magic classical quantity}
\end{equation}
Hence, Eq. (\ref{fundamental equation}) becomes 
\begin{eqnarray}
\underline{\rho}\mapsto\rho P(\lambda),\hspace{20mm}\nonumber\\
\partial_{q_i}\underline{S}\mapsto\partial_{q_i}S+\frac{\lambda}{2}\frac{\partial_{q_i}\rho}{\rho},\hspace {2mm}i=1,2,\hspace{0mm}\nonumber\\
\partial_{t}\underline{S}\mapsto\partial_tS+\frac{\lambda}{2}\frac{\partial_t\rho}{\rho}+\frac{g\lambda}{2}\partial_{q_1}\big(B\partial_{q_2}S\big)\nonumber\\
+\frac{g\lambda}{2}\partial_{q_2}\big(B\partial_{q_1}S\big).\hspace{0mm}
\label{fundamental equation classical quantity}
\end{eqnarray}

Let us see how the above set of equations change Eqs. (\ref{H-J equation classical quantity}) and (\ref{continuity equation classical quantity}). Imposing the first two equations of (\ref{fundamental equation classical quantity}) into Eq. (\ref{continuity equation classical quantity}) one has 
\begin{eqnarray}
\partial_t\rho+g\partial_{q_1}\big(\rho B\partial_{q_2}S\big)+g\partial_{q_2}\big(\rho B\partial_{q_1}S\big)\nonumber\\
+\frac{g\lambda}{2}\partial_{q_1}\big(B\partial_{q_2}\rho\big)+\frac{g\lambda}{2}\partial_{q_2}\big(B\partial_{q_1}\rho\big)=0. 
\label{FPE classical quantity}
\end{eqnarray}
Next, imposing the last two equations of Eq. (\ref{fundamental equation classical quantity}) into Eq. (\ref{H-J equation classical quantity}) one obtains, after arrangement, 
\begin{eqnarray}
\partial_tS+gB\partial_{q_1}S\partial_{q_2}S-g\lambda^2B\frac{\partial_{q_1}\partial_{q_2}R}{R}+\frac{g\lambda^2}{2}\Big(\partial_{q_1}B\frac{\partial_{q_2}R}{R}\Big)\nonumber\\
+\frac{\lambda}{2\rho}\Big(\partial_t\rho+g\partial_{q_1}\big(\rho B\partial_{q_2}S\big)+g\partial_{q_2}\big(\rho B\partial_{q_1}S\big)\hspace{20mm}\nonumber\\
+\frac{g\lambda}{2}\partial_{q_1}\big(B\partial_{q_2}\rho\big)+\frac{g\lambda}{2}\partial_{q_2}\big(B\partial_{q_1}\rho\big)\Big)=0,\hspace{5mm}\nonumber\\ 
\label{HJM classical quantity}
\end{eqnarray}
where $R\doteq\sqrt{\rho}$ and we have used Eq. (\ref{fluctuation decomposition}). Inserting Eq. (\ref{FPE classical quantity}), one thus obtains 
\begin{eqnarray}
\partial_tS+gB\partial_{q_1}S\partial_{q_2}S\hspace{40mm}\nonumber\\
-g\lambda^2B\frac{\partial_{q_1}\partial_{q_2}R}{R}+\frac{g\lambda^2}{2}\Big(\partial_{q_1}B\frac{\partial_{q_2}R}{R}\Big)=0. 
\label{HJM classical quantity}
\end{eqnarray}
We have thus pair of coupled equations (\ref{FPE classical quantity}) and (\ref{HJM classical quantity}) which are parameterized by the random variable $\lambda=\pm\hbar$. 

One can again see that $S(q,\hbar;t)=S(q,-\hbar;t)=S_Q(q;t)$ satisfies the same differential equation of (\ref{HJM classical quantity}) where $\lambda^2$ is replaced by $\hbar^2$. Hence, averaging over the fluctuations of $\lambda=\pm\hbar$ which is assumed to be equally probable, Eqs. (\ref{FPE classical quantity}) and (\ref{HJM classical quantity}) become 
\begin{eqnarray}
\partial_t\rho+g\partial_{q_1}\big(\rho B\partial_{q_2}S_Q\big)+g\partial_{q_2}\big(\rho B\partial_{q_1}S_Q\big)=0,\hspace{3mm}\nonumber\\ 
\partial_tS_Q+gB\partial_{q_1}S_Q\partial_{q_2}S_Q\hspace{40mm}\nonumber\\
-g\hbar^2B\frac{\partial_{q_1}\partial_{q_2}R}{R}+\frac{g\hbar^2}{2}\Big(\partial_{q_1}B\frac{\partial_{q_2}R}{R}\Big)=0. 
\label{Madelung equation classical quantity}
\end{eqnarray}

Recalling Eq. (\ref{Schroedinger wave function}) that $\Psi_Q=R\exp(iS_Q/\hbar)$, the above pair of coupled equations can be written into the Schr\"odinger equation $i\hbar\partial_t\Psi_Q=\hat{H}\Psi_Q$ with quantum Hamiltonian given by 
\begin{equation}
{\hat H}=\frac{g}{2}\big(B(q_1){\hat p}_1+{\hat p}_1B(q_1)\big){\hat p}_2. 
\end{equation}
Hence, comparing the above equation with Eq. (\ref{classical Hamiltonian classical quantity}), we have the following quantization mapping in the context of measurement: 
\begin{eqnarray}
\underline{p}_2\mapsto\hat{p}_2,\hspace{20mm}\nonumber\\
B(q_1)\underline{p}_1\mapsto\frac{1}{2}\big(B(q_1){\hat p}_1+{\hat p}_1B(q_1). 
\end{eqnarray}


\begin{thebibliography}{10}
\bibitem{constrained motion} Let us note that the classical velocity field will depend explicitly on the HPF only if the corresponding Lagrangian is not singular so that the classical velocity is expressible in term of classical momentum: $\dot{q}=\dot{q}(q,\underline{p};t)=\dot{q}(q,\partial_q\underline{S};t)$. Otherwise one has a constrained dynamics.
\end{thebibliography}
\end{document}